\documentclass[conference]{IEEEtran}
\IEEEoverridecommandlockouts

\usepackage{cite}
\usepackage{amsmath,amssymb,amsfonts}
\usepackage{graphicx}
\usepackage{textcomp}
\usepackage{xcolor}
\usepackage{algorithm}
\usepackage{algpseudocode}
\usepackage{float}
\usepackage{booktabs} 
\usepackage{setspace}

\usepackage{tabularx}

\usepackage{siunitx} 
\usepackage{subcaption}
\usepackage{url}

\def\BibTeX{{\rm B\kern-.05em{\sc i\kern-.025em b}\kern-.08em
    T\kern-.1667em\lower.7ex\hbox{E}\kern-.125emX}}
\begin{document}

\setstretch{.95}
\title{\Huge{A Versatile and Programmable UAV Platform for Radio Access Network and End-to-End Cellular Measurements} 
}



\author{
\IEEEauthorblockN{Sherwan Jalal Abdullah\IEEEauthorrefmark{1},  Sravan Reddy Chintareddy\IEEEauthorrefmark{1}, Victor S. Frost\IEEEauthorrefmark{1}, Shawn Keshmiri\IEEEauthorrefmark{2}, Morteza Hashemi\IEEEauthorrefmark{1}}
\IEEEauthorblockA{\IEEEauthorrefmark{1}Department of Electrical Engineering and Computer Science, University of Kansas} \IEEEauthorblockA{\IEEEauthorrefmark{2}Department of Aerospace Engineering, University of Kansas}}
\maketitle

\begin{abstract}


In this work, we develop a measurement platform to capture mobile network performance metrics including coverage and quality of service in regions where conventional coverage testing approaches are frequently time-intensive, labor-demanding, and occasionally hazardous. Traditionally, crowd-sourcing methods are used to collect cellular network performance metrics. However, these approaches are inadequate in rural areas due to low-density population, and difficult terrain. The platform described here is a UAV-based and is designed to investigate the mobile network performance through aerial operations and gather Radio Access Network (RAN) signal alongside end-to-end network performance metrics.
Our platform gathers metrics through the integration of an onboard computation unit and commercial off-the-shelf cellular modem. The gathered data are subsequently analyzed and displayed using geospatial mapping utilities and statistical techniques to deliver key observations on cellular network performance.
Experimental results showed that the received signal power improves at higher altitudes due to enhanced line-of-sight (LoS) conditions as expected. However, the signal quality degrades as a result of increased interference from neighboring cells.
The analysis reveals that for most of the geographic area covered in the initial experiments the system maintained acceptable signal quality, with adequate throughput performance for both uplink and downlink communications, while maintaining satisfactory round-trip time characteristics. Notably, the experiment showed that a strong radio signal metric for a given cell does not necessarily translate to consistent spatial coverage across the tested region.

\end{abstract}


\section{introduction}
\label{sec:intro}

Rural areas increasingly depend on cellular connectivity for critical applications, including precision agriculture, telemedicine, remote education, and Internet of Things deployments~\cite{exp_eval}. These applications require reliable cellular performance data to ensure quality of service for users across diverse rural landscapes. However, network characterization in rural environments faces significant challenges due to geographical constraints and sparse infrastructure deployment.

Measuring cellular network performance in rural areas presents significant technical challenges that limit effective network characterization. Traditional drive testing encounters substantial limitations due to rough terrain, sparse road infrastructure, and vast agricultural areas that make systematic ground-based measurements impractical. The geographical constraints in rural environments often result in incomplete coverage maps, leaving critical service areas unmeasured. Crowd-sourced~\cite{crowd_source_limit} measurement platforms like CellMapper~\cite{cellmapper} and OpenSignal~\cite{opensignal} provide inadequate rural coverage due to low population density, creating substantial data gaps that affect network planning and optimization decisions.

\begin{figure}
    \centering
    \includegraphics[width=1\linewidth]{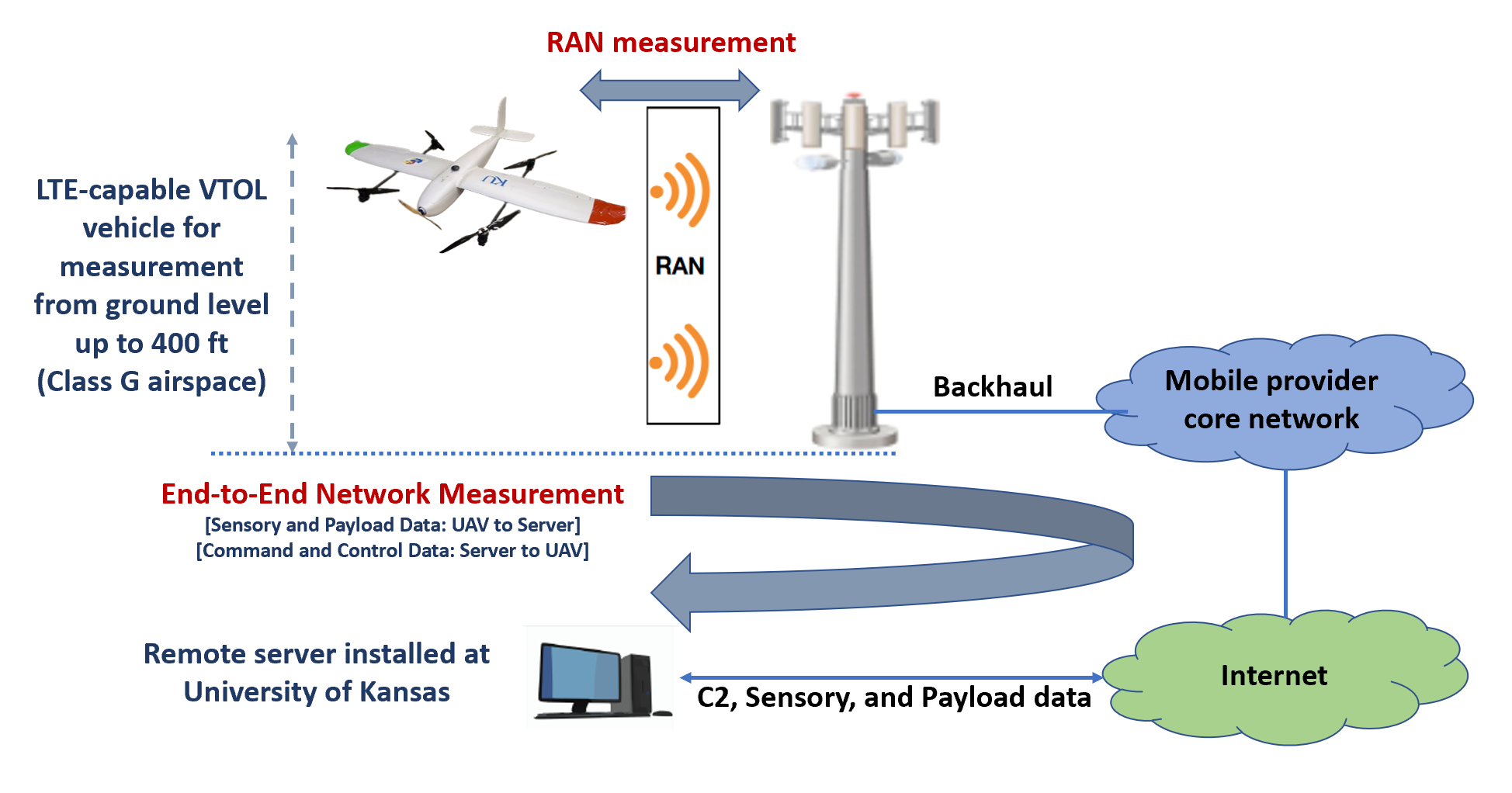}
    \caption{Overall system model and measurement scenario using UAV that can hover above the ground level (to mimic ground-level cellular devices) and fly up to an altitude of 400 ft., collecting Radio Access Network (RAN) and end-to-end network measurement data.}
    \label{fig:NSF_UAV}
\end{figure} 


Accurate cellular network performance metrics are essential for coverage tuning, identifying coverage gaps, and guiding network expansion in underserved rural regions. In such environments, traditional ground-based measurement campaigns are often hindered by challenging terrain and limited accessibility~\cite{crowd_source_limit,chintareddy2023collaborative,chintareddy2025federated}. Unmanned Aerial Vehicles (UAVs) provide an effective alternative that enables systematic three-dimensional data collection across diverse rural landscapes. 

Previous UAV-based cellular measurement studies~\cite{sae2019public,pathloss_measure,UL_IOI} have employed different approaches, each with significant limitations. Commercial measurement solutions rely on proprietary software platforms such as InfoVista TEMS~\cite{tems}, Rohde and Schwarz QualiPoc~\cite{qualipoc}, and Keysight Nemo Outdoor~\cite{nemo}. These tools create substantial barriers through expensive licensing requirements and closed architectures that prevent customization. Although feature-rich, these commercial solutions limit accessibility and reproducibility due to their proprietary nature and high costs.

Alternative approaches using Software-Defined Radio (SDR) systems~\cite{nekrasov2019evaluating,mobile_nw,eval_aerial} offer more accessible measurement capabilities but suffer from fundamental limitations in functionality and scope. SDR-based measurements typically capture only basic signal strength indicators like Reference Signal Received Power (RSRP) without providing essential network context such as cell identification, neighboring cell information, or higher-layer performance metrics. Recent UAV-mounted SDR investigations~\cite{LTE_AERPAW, volumetric} have focused primarily on single base station scenarios using private network testbeds, limiting their applicability to real-world interference-rich environments where multiple nodes compete for spectrum resources. 

Effective cellular network measurement requires frameworks that capture multiple layers of network behavior. Essential radio frequency indicators include Received Signal Strength Indicator (RSSI) for total received power assessment, (RSRP) for LTE reference signal strength evaluation, Reference Signal Received Quality (RSRQ) for signal quality measurement considering interference and noise, and Signal-to-Interference-plus-Noise-Ratio (SINR) for data rate prediction~\cite{tech_feas}. These serving cell metrics must be analyzed alongside neighboring cell information to understand interference patterns and handover behavior, particularly at elevated altitudes where line-of-sight conditions increase cell visibility. Complete network characterization also requires end-to-end performance evaluation, including latency measurements and bidirectional throughput testing to assess actual user experience under operational conditions. 

As illustrated in Figure \ref{fig:NSF_UAV}, we present a UAV-mounted cellular measurement system that combines commercial cellular modems with embedded computing platforms to address the above-mentioned limitations. Our approach provides an open-source implementation and customization capabilities while avoiding licensing restrictions. The system captures both radio layer metrics and end-to-end performance data, overcoming the cost and capability limitations of existing approaches. The platform enables systematic data collection across multiple altitudes and terrain types, enabling better assessment of rural network performance. To summarize, our main contributions are as follows:
\begin{itemize}
    \item We design and implement a UAV-mounted cellular measurement system that integrates commercial cellular modems with embedded computing platforms to capture radio access network parameters (RSRP, RSRQ, RSSI, SINR) and end-to-end performance metrics (latency, uplink and downlink throughput).
    \item We provide a measurement framework that enables data collection across multiple altitudes and geographical regions where terrestrial measurement approaches are impractical. 
    \item We create an open-source measurement platform that provides publicly accessible tools and methodologies for cellular network evaluation, facilitating collaborative research and broader access to advanced measurement capabilities. 
\end{itemize}
This paper is organized as follows. In Section \ref{sec:exsetup}, we present the measurement configuration, including hardware and software components. Section \ref{sec:na} presents comprehensive measurement results to demonstrate different capabilities of our developed measurement platform. Section \ref{sec:conclusion} concludes the paper and provides suggestions for future research.

\section{Experimental Setup}
\label{sec:exsetup}


In this section, we discuss the design and implementation of our UAV-based measurement system, including hardware configuration, software implementation, and the collected performance metrics to evaluate the network performance within the chosen test environment.

\subsection{Hardware Configuration
}
\begin{figure}
    \centering
   \includegraphics[width=1\linewidth]{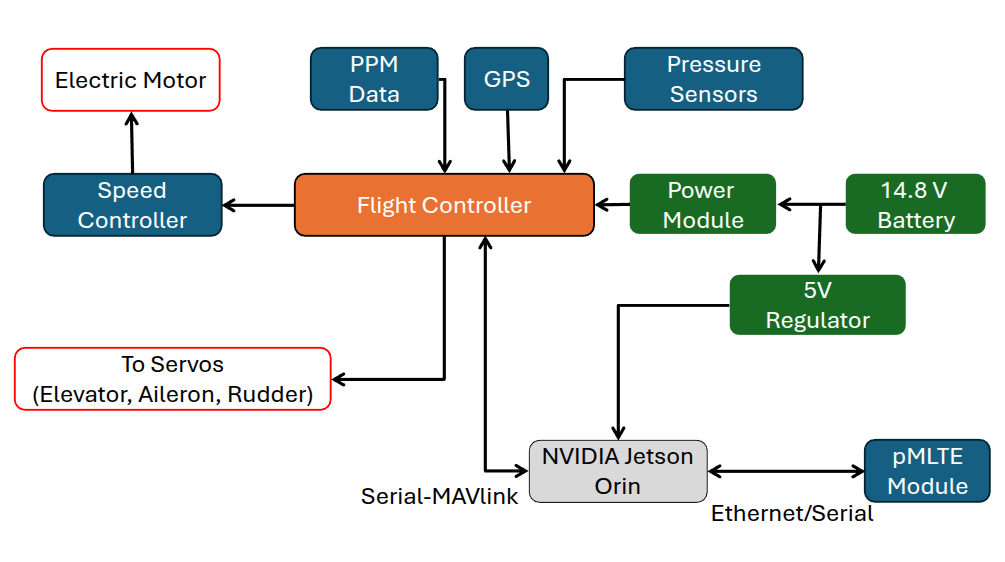}
    \caption{Internal system components of the UAV equipped with a flight controller, NVIDIA Orin processor, and pMLTE cellular modem. All components are fully integrated to provide a unified framework for cellular measurements.}
    \label{fig:hw}
\end{figure}

As shown in Fig.~\ref{fig:NSF_UAV}, we developed a UAV-based platform that is integrated with an in-house developed Automatic Flight System (AFS $5.0$). The hardware architecture of the proposed system is illustrated in Figure~\ref{fig:hw}, detailing the integrated avionics onboard the UAV. At the core of the system is the NVIDIA Jetson Orin~\cite{jetson}, which serves as the primary processing unit. This module comes pre-installed with the NVIDIA JetPack SDK and operates on a Linux-based operating system. The Robot Operating System (ROS) framework~\cite{ros} is deployed on the Orin to manage and execute all UAV software functionalities. The Jetson Orin interfaces directly with the flight controller, which acts as the main data acquisition unit by collecting real-time information from the onboard inertial measurement unit (IMU), GPS, and pressure sensors. Additionally, the flight controller manages servo actuation through the Orin's offboard control interface, with control signals transmitted via MAVLink protocol messages over a USB-serial connection.

To capture the cellular network performance metrics, the system is integrated with a commercial off-the-shelf (COTS) cellular modem, specifically the Microhard pMLTE~\cite{pmlte}. This cellular modem integrates a high-power Digital Data Link (DDL) with a robust LTE interface in a compact and durable form factor. The DDL subsystem operates in $2.4$ GHz band and includes several features such as $2$×$2$ MIMO, Maximal Ratio Combining (MRC), Maximal Likelihood (ML) decoding, and Low-Density Parity Check (LDPC) coding, ensuring robust wireless performance.  The LTE interface provides compatibility with globally deployed LTE networks and $3$G/HSPA fallback capability. The module supports LTE FDD with a maximum $150$ Mbps downlink/$50$ Mbps uplink and LTE TDD with a maximum $130$ Mbps downlink/$30$ Mbps uplink, utilizing standard Nano (4FF) SIM cards for connectivity.  It supports adjustable high-power transmission up to $1$ W total, with enhanced receiver sensitivity and MIMO antenna alignment. In addition to wireless capabilities, the pMLTE provides a configurable firewall with ACL-based security, port forwarding, and supports a wide range of network protocols, including TCP/UDP, TFTP, ARP, ICMP, DHCP, HTTPS, SSH, SNMP, FTP, DNS, and Serial-over-IP. Additionally, the pMLTE supports dual $10$/$100$ Ethernet (LAN/WAN) ports, dual RS-$232$ TTL serial/console interfaces, and an integrated GPS receiver. 

\subsection{Software Implementation and collected performance metrics
}
The pMLTE modem’s built-in WebGUI interface provided basic data collection functionality but introduced multiple limitations for UAV-based LTE performance measurements. It was constrained to capturing only RSSI values among the radio access metrics and enforced a fixed sampling rate of $10$ seconds, which proved insufficient for tracking rapid variations in signal characteristics during continuous aerial movement. Additionally, the WebGUI could not measure end-to-end performance metrics such as latency and throughput, which are essential for evaluating actual service delivery. To overcome these constraints, we developed a custom data logger application that is deployed on the Jetson Orin onboard computer. The data logger application communicates with the pMLTE modem through the Ethernet link and enables high-resolution data acquisition.  The data logger operates at a $1$-second sampling interval, significantly enhancing temporal granularity, and records each measurement with precise GPS coordinates and timestamps to facilitate synchronized spatial and temporal analysis.



Furthermore, the logging software retrieves a set of radio parameters, including RSRP, RSRQ, RSSI, and SINR, along with network-layer performance metrics such as latency, uplink (UL), and downlink (DL) throughput.  Radio frequency evaluation encompasses RSRP as the primary signal strength indicator, RSSI quantifying total received power including interference and noise elements, RSRQ characterizing radio link quality parameters, and SINR measuring signal-to-interference-plus-noise-ratio that directly influence achievable transmission rates.
Network infrastructure analysis incorporates the Physical Cell ID (PCI) for Radio Access Network and Cell ID for higher network layer identifications. The Location Area Code (LAC) provides regional network architecture insights, while neighboring cell measurements enable handover target analysis for service reliability assessment.

End-to-end performance characterization extends beyond radio frequency analysis to evaluate actual service delivery capabilities. Packet round-trip time measurements employ Nping~\cite{nping} 
which is an open source tool for network packet generation, response analysis and response time measurement. while bidirectional throughput evaluation uses iPerf$3$~\cite{iperf} which is a tool for active measurements of the maximum achievable bandwidth on IP networks. The remote server endpoint approach mitigates content delivery network variability that could potentially compromise measurement accuracy.

\subsection{Test Environment}

Field experiments were conducted in the rural environment near Clinton Lake, Lawrence, Kansas, which presents representative coverage assessment challenges, including varied terrain conditions and sparse infrastructure deployment. 
This integrated measurement approach systematically addresses the disconnect between theoretical coverage predictions and practical user experience metrics. The framework facilitates the analysis of coverage distribution patterns and service quality characteristics previously unattainable through traditional measurement methodologies. The platform presented here provides a reproducible methodology for rural network evaluation that informs both research investigations and practical network planning strategies in complex deployment environments.

\section{Experimental Results and Analysis}
\label{sec:na}

This section examines the experimental results of UAV-based network measurements recorded within the rural evaluation environment. The measurement data were thoroughly processed, analyzed, and visualized to provide a detailed assessment of LTE radio frequency characteristics and end-to-end service performance metrics across various flight positions and altitude levels.

\subsection{Spatial Coverage Analysis}

\begin{figure}
    \centering
    \includegraphics[width=1\linewidth]{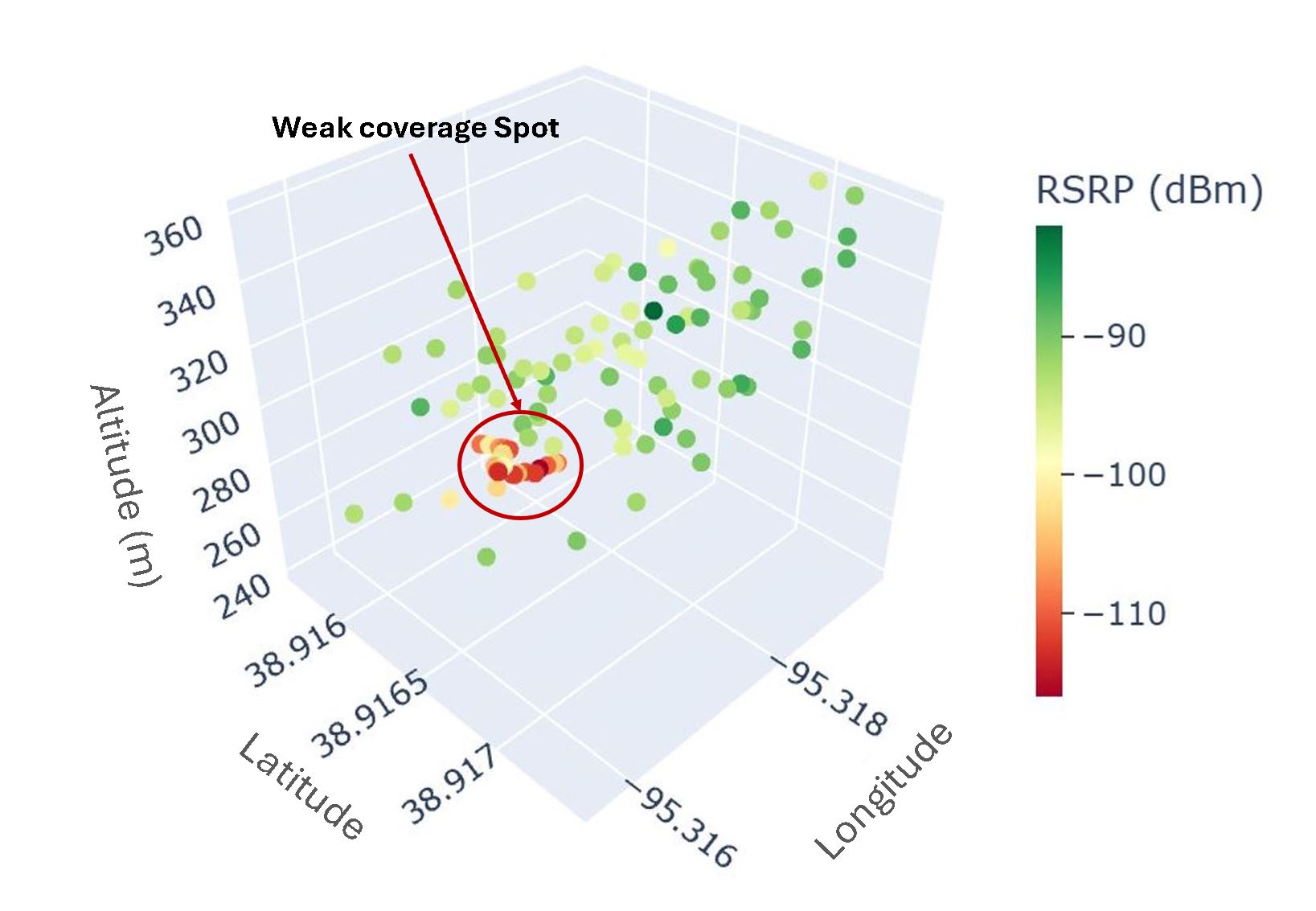}
    \caption{$3$D visualization of RSRP measurements in the test area showing signal strength distribution across GPS coordinates and altitude above sea level.}
    \label{fig:3D_RSRP_trajectory_map}
\end{figure}

To demonstrate the spatial and vertical variability of LTE coverage, geospatial visualizations were generated in both $2$D and $3$D formats. These representations illustrate the distribution and behavior of the performance metrics (RSRP, RSRQ, RSSI, and SINR) across various positions and altitudes throughout the flight trajectories. This visualization approach provides a general overview of network performance by identifying the coverage gaps, interference patterns, and signal quality variations. Furthermore, the $3$D visualization additionally reveals how signal strength and quality change with altitude. Figure~\ref{fig:3D_RSRP_trajectory_map} presents the $3$D RSRP distribution across the test environment, demonstrating the spatial coverage characteristics and signal strength variations throughout the measurement campaign. The visualization clearly indicates a weak spot at lower altitudes, which could have been caused by the geographical terrain that naturally obstructs the signal.


\begin{figure}
    \centering
    \includegraphics[width=1\linewidth]{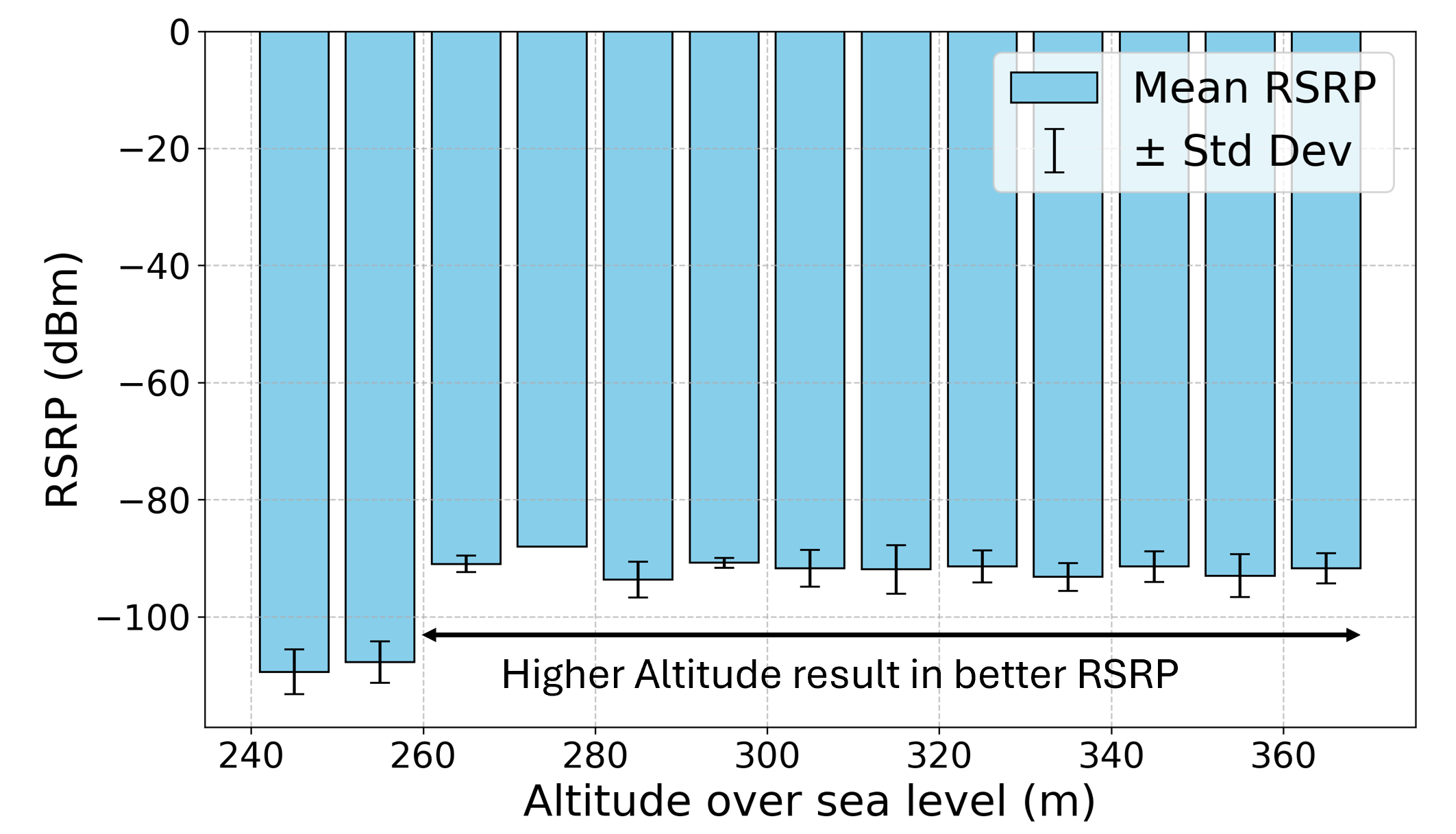}
    \caption{Altitude-dependent RSRP (dBm) performance analysis showing improved signal reception at higher altitudes.}
    \label{fig:rsrp_vs_altitude}
\end{figure}

Given that experimental measurements were conducted across multiple altitude levels, an altitude-based analysis was performed to evaluate the performance metrics. Figure~\ref{fig:rsrp_vs_altitude} shows that the measured RSRP values increase for higher altitudes, indicating enhanced received signal power characteristics. This enhancement is primarily attributed to improved Line-of-Sight (LoS) propagation conditions between the UAV platform and base station infrastructure, which minimize signal attenuation effects caused by terrestrial obstacles, including terrain features and vegetation coverage. We also observed a similar trend with RSSI measurements, which are not included in this paper due to space limitations. Conversely, while RSRP and RSSI metrics demonstrate improvement with increasing altitude, RSRQ and SINR parameters exhibit performance degradation. This deterioration results from enhanced neighboring cell visibility at elevated positions, which increases interference levels during reference signal reception processes. 
This altitude-dependent trade-off between signal power and signal quality represents a fundamental challenge for UAV communications, where improved line-of-sight conditions simultaneously enhance desired signal reception while increasing interference from neighboring cells.

 \subsection{Radio Signal Characteristics Analysis}

\begin{figure}
    \centering
    \includegraphics[width=1\linewidth]{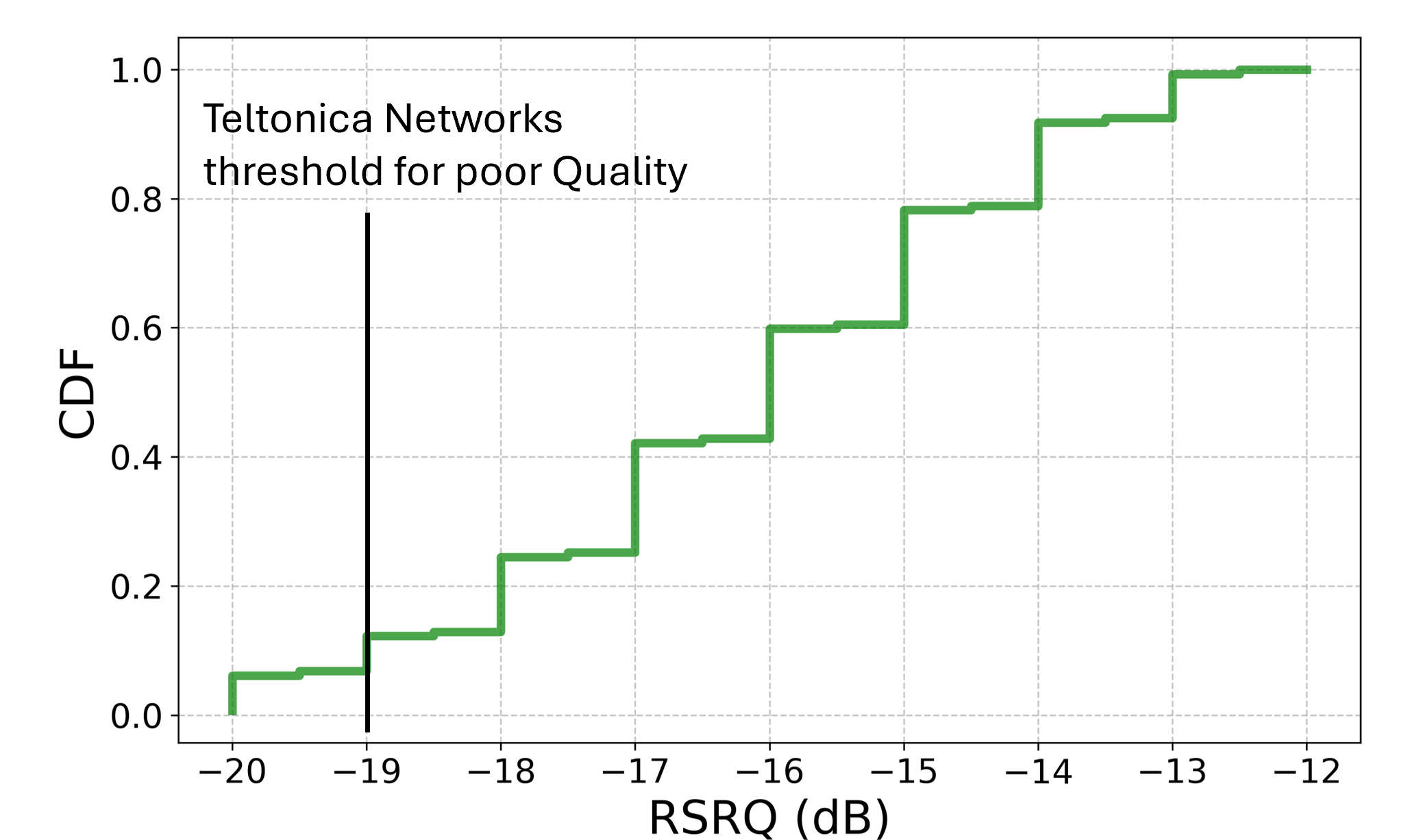}
    \caption{CDF of RSRQ measurement demonstrating signal quality distribution with Teltonica Networks threshold for poor-quality signal identification at $-19$ dB.}
    \label{fig:CDF_RSRQ}
\end{figure}

While spatial visualization techniques provide effective geographic representation of radio signal characteristics, they do not reveal detailed, quantified distribution properties of acquired measurements. Consequently, a quantitative evaluation was performed to analyze the performance attributes of LTE signal metrics across the test deployment area. Cumulative Distribution Function (CDF) analyses were employed to characterize the value-based distribution of essential performance metrics. 
These analytical results provide detailed signal quality characterization, enabling quantification of the area proportion experiencing specific Radio Access Network parameter thresholds. Figure \ref{fig:CDF_RSRQ} illustrates the RSRQ value distribution measured in dB throughout the tested region, presenting the statistical distribution of signal quality performance across the experimental environment. Since standardized thresholds for RSRQ quality classification do not exist, this study employs criteria proposed by Teltonika Networks~\cite{teltonika}, utilizing $-19$ dB as the poor quality threshold. According to this standard, the CDF evaluation reveals that approximately $15\%$ of the assessed area experiences poor quality radio signal conditions.

\begin{figure}
    \centering
    \includegraphics[width=1\linewidth]{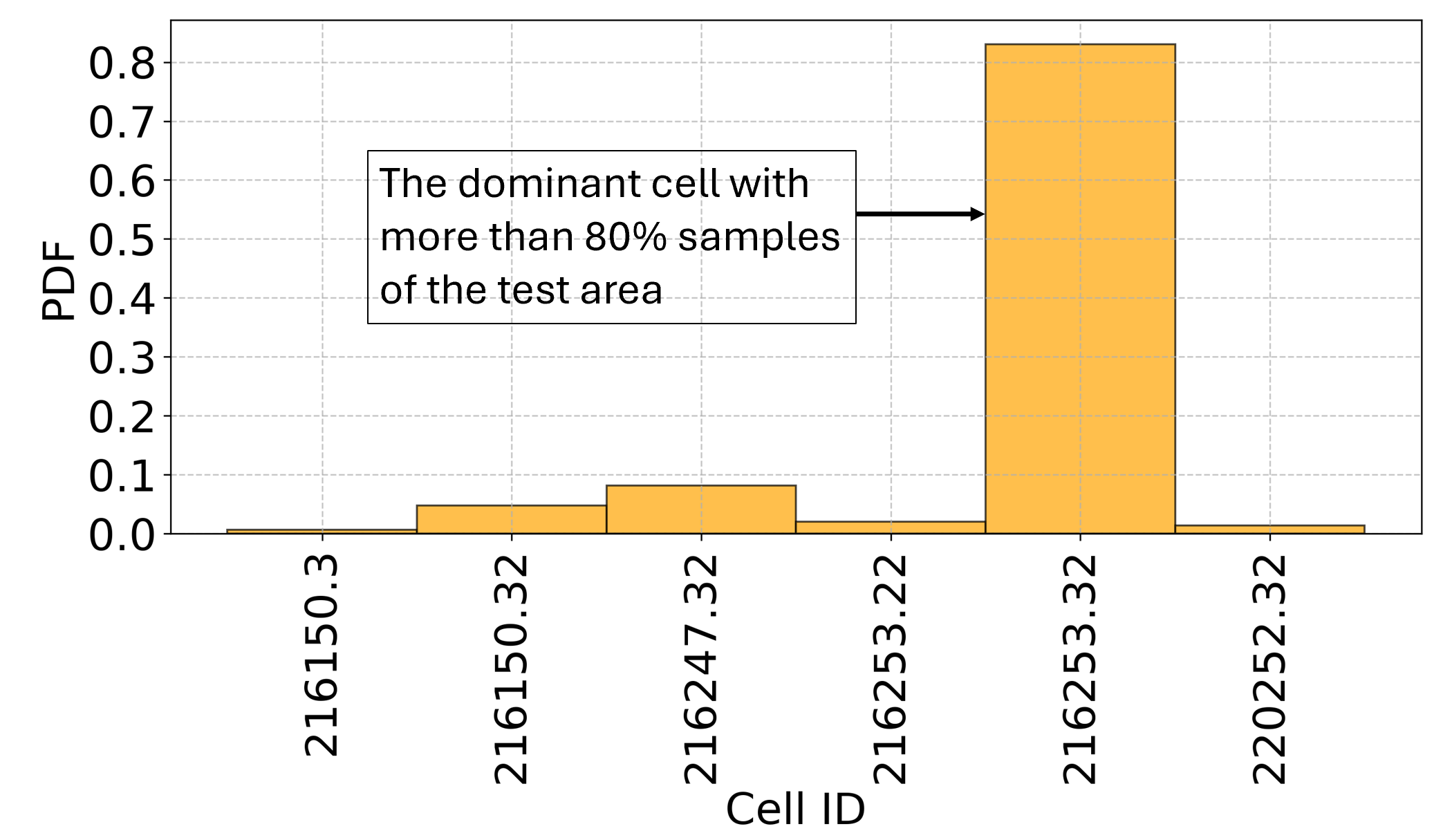}
    \caption{PDF of serving cell coverage dominance, with one cell providing majority coverage across the measurement area.}
    \label{fig:PDF_CellID}
\end{figure}

\begin{figure}
    \centering
    \includegraphics[width=1\linewidth]{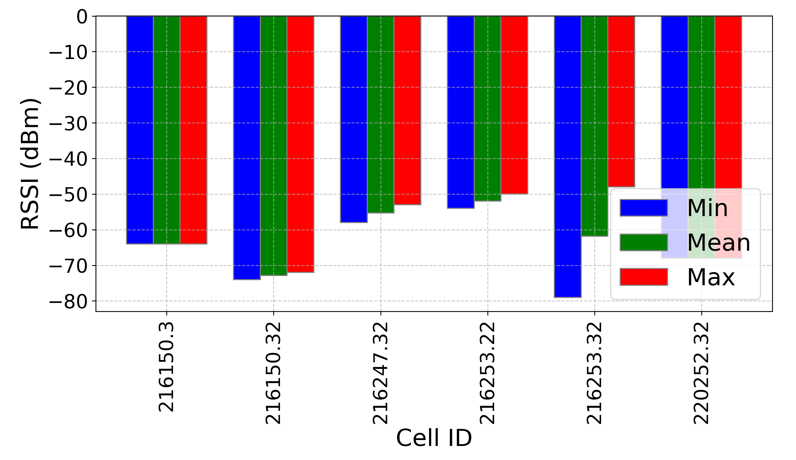}
    \caption{Comparative RSSI (dBm) performance showing signal strength bounds and average values for each Cell ID during the measurement flight.}
    \label{fig:stats_RSSI}
\end{figure}

Since cellular network coverage is provided by multiple cells with varying contributions to the received service, understanding individual cellular performance becomes crucial for network assessment. To gain insight into individual cellular contributions to service delivery within the test deployment, cell-based performance evaluation was conducted. This methodology enables individual cellular assessment regarding coverage dominance and signal quality parameters. This knowledge support strategic network optimization and focused improvement. The evaluation results demonstrated in Figures~\ref{fig:PDF_CellID} and~\ref{fig:stats_RSSI} present the distribution characteristics and statistical performance of radio metrics across different cells.
\begin{figure}
    \centering
    \includegraphics[width=1\linewidth]{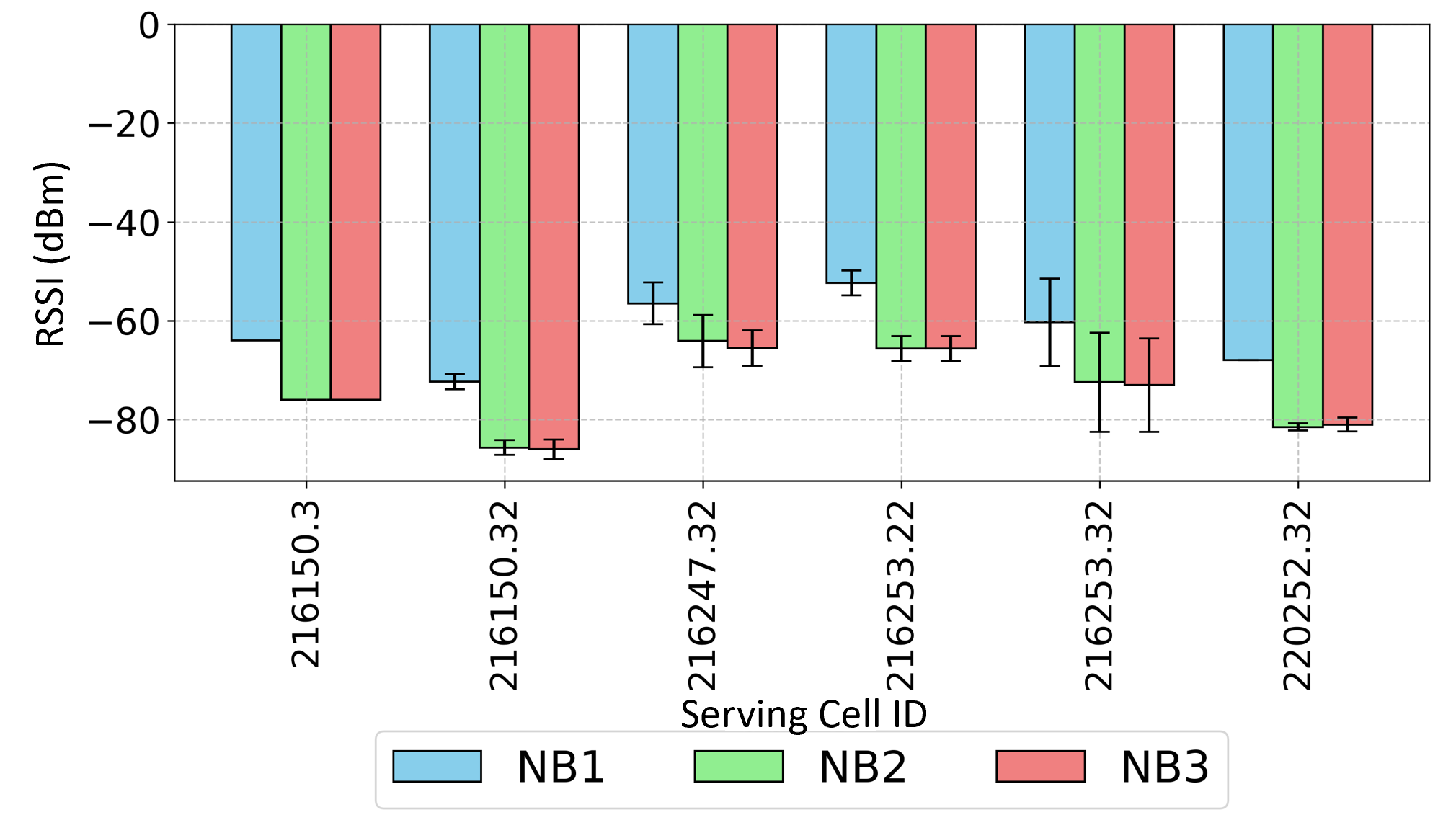}
    \caption{Statistical analysis of neighbor cells RSSI measurements showing mean values (shaded regions) and standard deviation (line boundaries).}
    \label{fig:nb_rssi}
\end{figure}
However, examination of these findings reveals that the signal parameters associated with particular cells do not necessarily indicate a good overall coverage quality within the area, since these cells may not contribute with high number of samples that were collected from the tested area. 
This finding underscores the essential requirement to evaluate both signal quality parameters and cellular coverage in comprehensive network performance analysis.

The data logger software captures samples of LTE radio metrics from neighboring cells, which serve as critical handover candidates during serving cell signal degradation. These neighboring cells RAN measurements are fundamental for assessing network reliability and maintaining seamless connectivity across the coverage area. The log of neighboring cell data enables an evaluation of the radio frequency environment beyond the primary serving cell infrastructure. Figure \ref{fig:nb_rssi} demonstrates the statistical characterization of neighboring cell performance, presenting mean values alongside standard deviation metrics for radio parameters throughout the experimental evaluation area. The results show that at least one neighbor cell maintains good coverage status, enabling a smooth handover process, preventing service drops when the serving cell coverage degrades.




\begin{figure}
    \centering
    \includegraphics[width=1\linewidth]{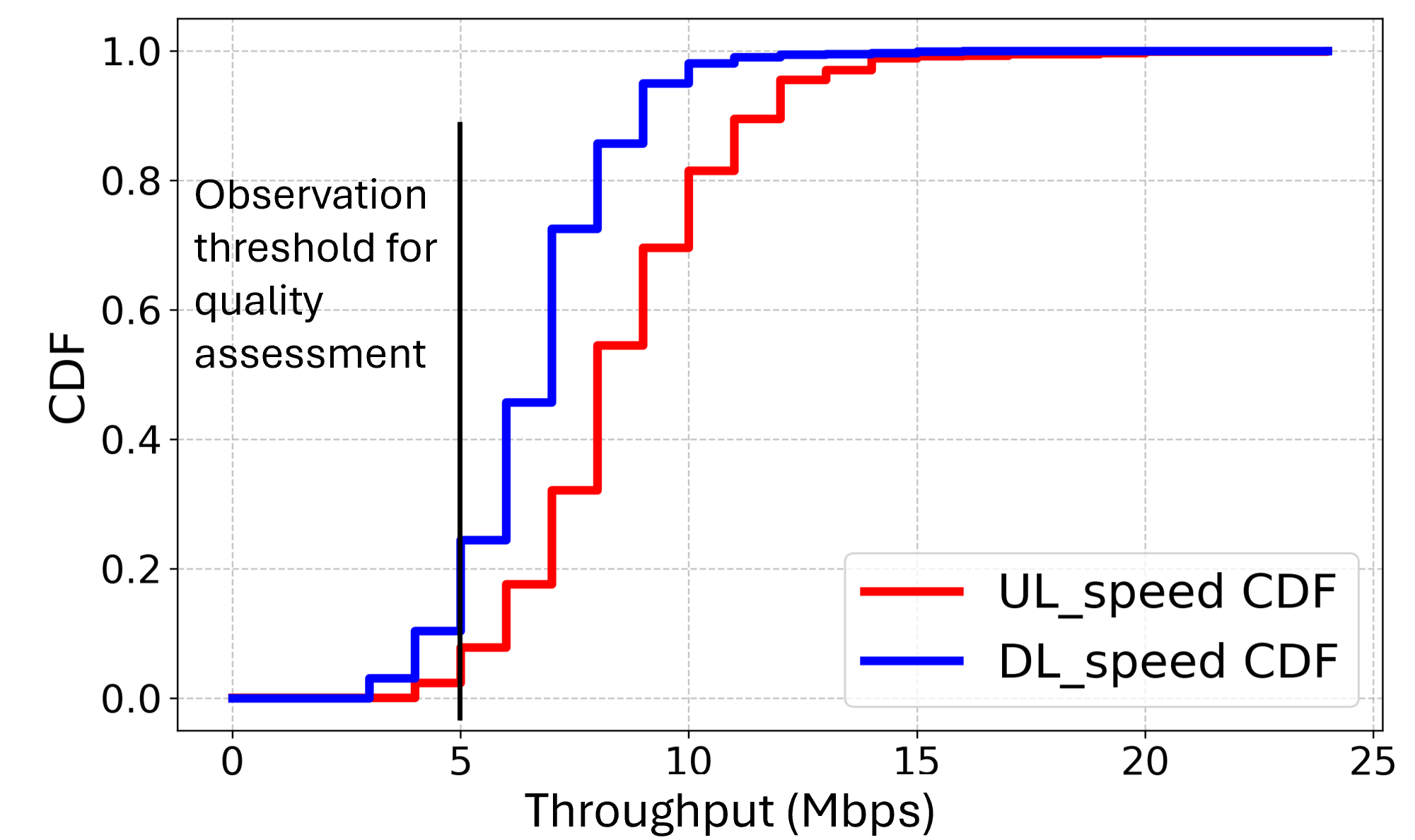}
    \caption{CDF of downlink and uplink speed measurements showing network performance characteristics with observation threshold at $5$ Mbps for quality assessment.}
    \label{fig:CDF_Mbps}
\end{figure}

\begin{figure}
    \centering
    \includegraphics[width=1\linewidth]{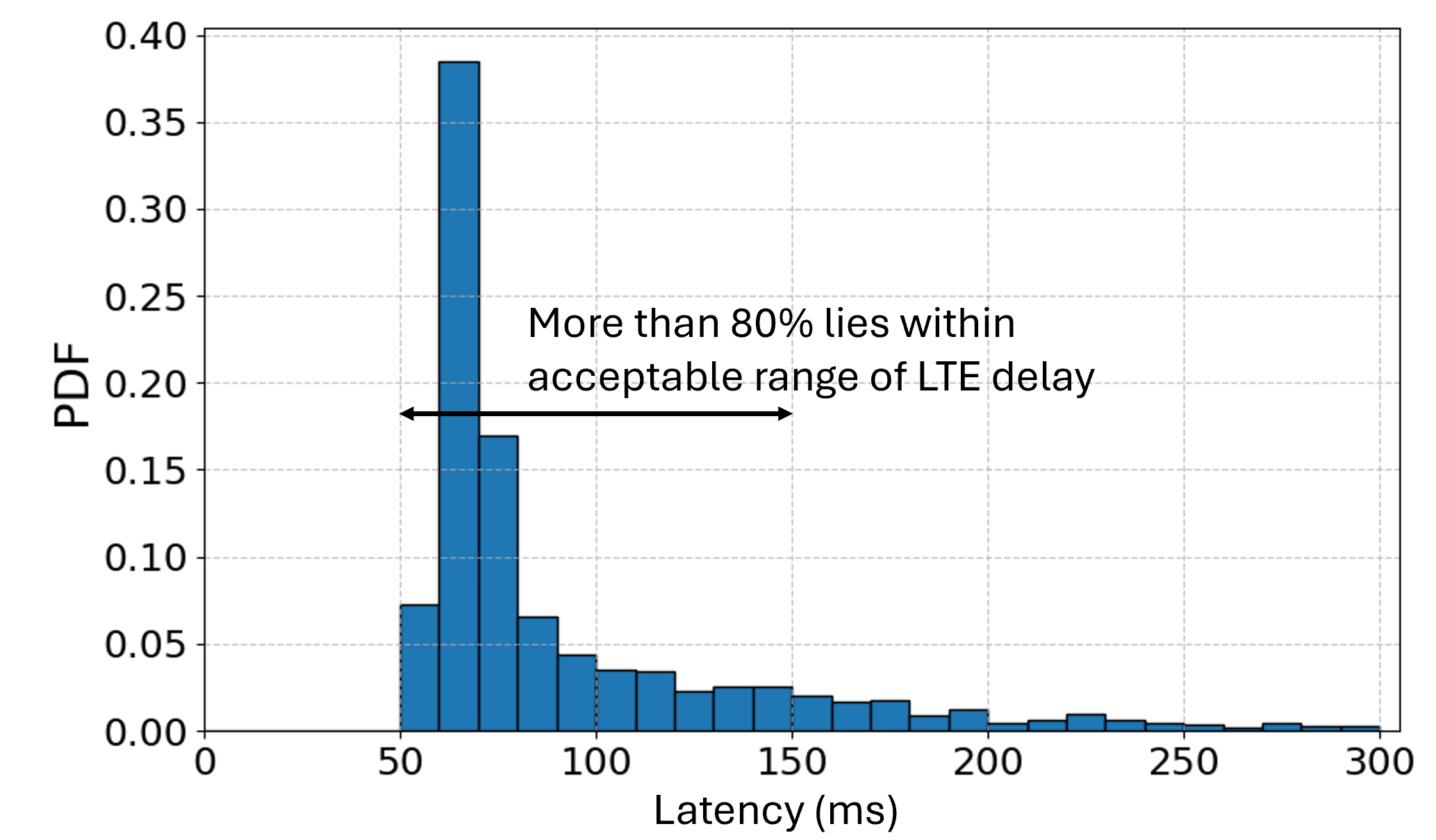}
    \caption{PDF analysis of RTT (ms) distribution showing that more than $80\%$ of measurements fall within $150$ ms.}
    \label{fig:PDF_Delay}
\end{figure}

\subsection{ End-to-End Service Performance Analysis}

While radio link metrics provide essential wireless connection characterization in mobile networks, they do not fully represent application-level performance experienced by end users. Analysis of application performance requires systematic evaluation of end-to-end network metrics, including packet round-trip time (RTT) and bidirectional throughput measurements that directly impact user applications. The logger software included in our framework connect a dedicated remote server infrastructure deployed at the University of Kansas to prevent measurement variations introduced by automated server selection processes utilized in conventional tools like OOKLA Speedtest~\cite{ookla}, and allows for a reliable comparison and comparative analysis among tests conducted at any time.
Figures~\ref{fig:CDF_Mbps} and~\ref{fig:PDF_Delay} provide statistical characterization of throughput performance and RTT distributions, respectively, demonstrating application-level service quality delivered through the LTE deployment. Performance analysis indicates that more than $90\%$ of the coverage region supports application throughput exceeding $5$ Mbps for both upload and download operations, while RTT performance between the pMLTE device and the dedicated server maintains levels below $150$ms for more than $80\%$ of the measurement locations.

\section{Conclusion and Future Works}
\label{sec:conclusion}

This work introduces a UAV-based measurement platform for cellular network characterization in rural environments where traditional drive tests face significant challenges. The system integrates UAV capabilities with commercial cellular modems and custom data logging software to enable systematic collection of both Radio Access Network parameters and end-to-end performance metrics. Experimental results reveal altitude-dependent variations: signal power improves at higher altitudes due to enhanced line-of-sight conditions, whereas signal quality decreases because of increased interference from neighboring cells. The analysis shows that $85\%$ of the tested area achieves acceptable signal quality, with over $90\%$ supporting throughput greater than $5$ Mbps in both uplink and downlink, and more than $80\%$ of RTT measurements remained below $150$ ms. Furthermore, good radio signal performance for a cell does not always imply a good spatial coverage in the tested area. The open-source platform and associated measurement protocols developed here establishes a replicable methodology bridging theoretical coverage predictions and practical deployments, offering researchers accessible tools for
rural network evaluation.
 
Future work will focus on leveraging these measurements to build extrapolation models for predicting terrestrial coverage and improving rural network planning strategies.

\section*{Acknowledgment}
The material is supported, in part, by the NSF grants 1955561, 2212565, 2323189, and 2434113. Any opinions, findings, conclusions, or recommendations expressed in this material are those of the author(s) and do not necessarily reflect the views of NSF.

\bibliographystyle{IEEEtran}
\bibliography{ref}

\end{document}